\documentstyle[12pt]{article}
\begin{document}
\vskip 4 cm
\begin{center}
\Large{\bf TO QUANTIZE OR NOT TO QUANTIZE GRAVITY ?}
\end{center}
\vskip 3 cm
\begin{center}
{\bf AFSAR ABBAS} \\
Institute of Physics, Bhubaneswar-751005, India \\
(e-mail : afsar@iopb.res.in)
\end{center}
\vskip 20 mm  
\begin{centerline}
{\bf Abstract }
\end{centerline}
\vskip 3 mm

It is shown here that the Standard Model (SM) of particle physics
supports the view that gravity need not be quantized. 
It is shown that the SM gives a consistent description of the origin 
of the universe. It is guggested that the universe came into existence 
when the SM symmetry was broken spontaneously. This brings out
a complete and consistent model of the physical universe in the 
framework of a "semiclassical quantum gravity" theory. 

\newpage

To quantize or not to quantize gravity, is one of the outstanding 
questions facing physics today. Though there is still no compelling 
empirical evidence supporting quantization, the current theoretical 
prejudice is strongly in favour of the latter possibilty.
Here it shall however be shown that the most succesful model of 
particle physics, the so-called Standard Model ( SM ), 
supports the former view and infact gives a consistent and complete
basis for the " semiclassical quantum gravity " idea of Rosenfeld [1]. 

As the universe expands, it is predicted
that it undergoes a series of phase transitions [2]  during which
the appropriate symmetry breakes down in various stages until it
reaches the stage of the SM symmetry
$SU(3)_c$ $\otimes$ $SU(2)_{L}$
$\otimes$ $U(1)_{Y} $. After $ t \sim 10^{-10} $ seconds 
( at  $T\sim 10^{2}$ GeV ) the SM phase 
transition to $SU(3)_c$ $\otimes$ $U(1)_{em}$
through the Higgs Mechanism takes place.

The SM has been well studied in the laboratory. It is
the best tested model of particle physics [3]. It was however 
believed earlier that a weaknesses of the SM was that the 
electric charge was not quantized in it. 
However, it was as late as 1990 that this was shown to be wrong. Against
all expectations, the electric charge was shown to be fully and
consistently quantized in the SM [4]. 

It was shown by the author [4,5] that in SM 
$ SU(N_{C}) \otimes SU(2)_{L} \times U(1)_{Y} $
for $ N_{C} = 3 $,
spontaneous symmetry breaking by a Higgs isospin doublet of weak 
hypercharge
$ Y_{ \phi } $,
the isospin -1/2 component 
develops the vacuum
expectation value $ < \phi >_{0} $. This  fixes `b' in the electric charge
definition $ Q = T_{ 3 } + b Y $ to give

\begin{equation}
Q = T_{ 3 } + \frac { 1 } { 2 Y_{ \phi } }  Y
\end{equation}

where Y is the hypercharge for doublets and singlets for
a single generation.
For each generation renormalizability through triangular anomaly
cancellation and the requirement of the identity of L- and R-handed charges
in $ U(1)_{ em } $ one finds that all unknown hypercharges are
proportional to 
$ Y_{\phi}$. Hence correct charges (for 
$ N_{C} = 3 $ ) fall through as below [4-8]

\begin{eqnarray}
Q(u) = \frac{1}{2} ( 1 + \frac{1}{N_{C}} ) \\
Q(d) = \frac{1}{2} ( -1 + \frac{1}{N_{C}} ) \\
Q(e) = -1 \\
Q(\nu) = 0
\end{eqnarray}

Hence the electric charge is
quantized in SM. The complete structure of the SM as is, is required
to obtain this result on very general grounds. 

Clearly the $ U(1)_{em}$ symmetry which
arose due to spontaneous symmetry breaking due to a Higgs doublet
in the SM symmetry will be lost above $ T_{c}^{SM} $ whence
$ SU(2)_{L}$ $\otimes$ $ U(1)_{em}$ symmetry would be
restored. As is obvious, above $ T_{c}^{SM}$ all the fermions and gauge 
bosons becomes massless. Here I point out a new 
phenomenon arising from the restoring of the full SM symmetry .

Note that to start with the parameter b and Y in the definition
of electric charge were completely unknown. We could lay a handle on 'b'
entirely on the basis of the presence of spontaneous symmetry breaking
and on ensuring that photon was massless $ b = \frac{1}{2 Y_{\phi}} $.
Above $ T_{c}^{SM} $ where the SM symmetry is restored there is
no spontaneous symmetry breaking and hence the parameter b is
completely undetermined. Together 'bY' could be any arbitrary number
whatsoever even an irrational number. Within
the framework of this model above $ T_{c}^{SM} $ we just cannot define 
electric charges at all. Hence the electric charge
loses any physical meaning all together.
There is no such thing as charge anymore.
The photon (which was a linear combination of $W^{0}$
and $ B_{\mu} $ after spontaneous symmetry breaking ), with it's
defining  vector characterteristic, 
does not exist either. So the conclusion is that 
there is no electrodynamics above $ T_{c}^{SM} $.
Maxwells equations of electrodynamics show that
light is an electromagnetic phenomenon. Hence above the SM phase
transition there was no light and no Maxwells equations. And surprisingly
we are led to conclude that there was no velocity of light c as well. One
knows that the velocity of light c is given as
 
\begin{equation}
c^2 = { 1 \over { \epsilon_0 \mu_0} }
\end{equation}

where $ \epsilon_0  $ and $ \mu_0 $ are permittivity and permeability
of the free space. These electromagnetic properties disappear
along with the electric charge and hence the velocity of light also
disappears above the SM phase transition.

The premise on which the theory of relativity is based is 
that c, the velocity of light is always the same, no matter from which
frame of reference it is measured. Of
fundamental significance is the invariant interval

\begin{equation}
s^2 = { ( c t ) }^2 - x^2 - y^2 - z^2
\end{equation}

Here the constant c provides us with a means of defining time in terms of
spatial separation and vice versa through l = c t. this enables one to
visualize time as the fourth dimension. Hence time gets defined due to
the constant c. Therefore when there were no c as in the early universe 
,there was no time as well. 
As the special theory of relativity depends upon c and
time, above the SM  breaking scale
there was no special theory of relativity. As the General Theory of
relativity also requires the concept of a physical space-time, it
collapses too above the SM breaking scale.
Hence the whole physical universe collapses above this scale. 

Therefore it was hypothesized that the Universe came into existence
when the SM symmetry was broken spontaneously [11].
Before it there was no 'time', no 'light', no maximum
velocity like 'c', no gravity or space-time structure on which objective
physical laws as we know it could exist [11].

In short we find that the SM of particle physics is not only good to
explain all known facts, it is also capable of giving a natural and
consistent description of the origin of the Universe [11].

So matter (which is quantizd) and the space-time ( and so gravity ) 
are bound together holistically. In this picture one cannot talk 
of matter without spacetime and neither can we talk of space-time 
without matter. They are two sides of the same coin and hence the 
Einstein's equation as per the SM should read as follows:

\begin{equation}
G_{\mu \nu} = 8 \pi G \langle \phi | T _{\mu \nu} | \phi \rangle
\end{equation}

This is the " semiclassical quantum gravity " approach of Rosenfeld [1].
However here $ \phi $ represents the complete matter degrees of freedom 
incorporated in the SM in terms of the three generations of matter 
particles plus the gauge particle of the SM 
( no other spurious model is needed [9,10] ).
Gravity, a property of the spacetime arises as a result of the quantum 
mechanical SM phase transtion and is not quantized. Hence as per the SM 
origin of the universe scenario presented here we get a complete and 
consistent picture of the origin and the existence of space-time and 
matter. The two are inseparably connected.

It seems that the reason that inspite of rigourous efforts 
we have not seen graviton so far is because as shown here it does not 
exist. This need not shock some modern adherents of quantization of 
gravity as many a scientists have been warning against an urge to quantize
gravity. Long ago Rosenfeld [1] had pointed out that empirical evidence 
and not logic forces us to quantize fields. He said that in the absence of 
such evidence the temptation to quantize fields must be resisted [1].
Indeed this is what the SM of particle physics is telling us.
And this is within the framework of the most successful model of particle 
physics. Not only can the SM of particle
physics provide a good description of all known particles, it also has the
capacity to give a consistent picture of the origin of the universe and
also of settling the issue of whether to quantize gravity or not.

\newpage

\vskip 2 cm
\begin{center}
{\bf REFERENCES }
m\end{center}

\vskip 1 cm

1. L. Rosenfeld, {\it Nucl. Phys. }, {\bf 40} (1963) 353

2. E. W. Kolb and M. S. Turner, " The Early Universe "
Edison Wesley, New York ( 1990 )

3. T. P. Cheng and L. F. Lee, " Gauge Theory of Elementary Particle
Physics ", Clarendon Press, Oxford ( 1988 )

4. A. Abbas, 
{\it Phys. Lett. }, {\bf B 238} ( 1990) 344

5. A. Abbas, 
{\it J. Phys. G. }, {\bf 16 } ( 1990 ) L163

6. A. Abbas, {\it Nuovo Cimento }, {\bf A 106} ( 1993 ) 985

7. A. Abbas, {\it Hadronic J.}, {\bf 15} ( 1992 ) 475

8. A. Abbas, {\it Ind. J. Phys.}, {\bf A 67} ( 1993 ) 541

9. A. Abbas,
{\it Physics Today } ( July 1999 )  p.81-82

10. A. Abbas,
{\it `On standard model Higgs and superstring theories'} 
"Particles, Strings and Cosmology PASCOS99", Ed K Cheung,
J F Gunion and S Mrenna, World Scientific, Singapore ( 2000) 123

11. A. Abbas,
" On the Origin of the Universe ",
{\it Spacetime \& Substance, Int Phys Journal}, {\bf 2 (7)} ( 2001 ) 49 ;
Website: spacetime.narod.ru/2-7-2001.html

\end{document}